\documentclass[10pt,twocolumn]{article}

\usepackage[T1]{fontenc}
\usepackage[utf8]{inputenc}
\usepackage{lmodern}
\usepackage{microtype}
\usepackage{textgreek}
\usepackage{authblk}

\usepackage[version=4]{mhchem}
\usepackage{verbatim}
\usepackage{makecell} 
\usepackage{siunitx} 
\usepackage[
  left=1.8cm,
  right=1.8cm,
  top=2.0cm,
  bottom=2.2cm,
  columnsep=0.45cm
]{geometry}

\usepackage{mathtools,amssymb}
\usepackage{booktabs}

\usepackage{float}

\newif\ifimgdraft
\imgdraftfalse

\ifimgdraft
  \usepackage[draft]{graphicx}
\else
  \usepackage{graphicx}
\fi

\usepackage{subcaption}
\usepackage{placeins}
\usepackage{dblfloatfix}

\usepackage[numbers,sort&compress]{natbib}
\usepackage{chapterbib}
\usepackage{newunicodechar}
\newunicodechar{×}{\times}
\usepackage[colorlinks,linkcolor=blue,citecolor=blue,urlcolor=blue]{hyperref}
\title{\bfseries
Electrolyte density, diffusivity and conductivity in graphene nanoconfinement predicted by separating interfacial from genuine confinement effects}

\author{%
Haoyuan Quan\textsuperscript{1},
Hanne S. Antila\textsuperscript{1},
Maximilian R. Becker\textsuperscript{1},
Philip R. Loche\textsuperscript{2},
and Roland R. Netz\textsuperscript{1,*}
\\
\normalsize\itshape
\textsuperscript{1}Freie Universit\"at Berlin, Department of Physics,
Institute for Theoretical Physics, Arnimallee 14,
14195 Berlin, Germany
\\
\textsuperscript{2}Laboratory of Computational Science and Modeling,
Institut des Mat\'eriaux, \'Ecole Polytechnique F\'ed\'erale de Lausanne,
1015 Lausanne, Switzerland
}

\date{}
\begin{document}


\twocolumn[
  \maketitle

    \begin{abstract}
    Confined aqueous electrolytes exhibit strong deviations from bulk behaviour, but it remains unclear which changes arise from genuine confinement-induced modification and which merely reflect the influence of interfaces. Here, we introduce an interfacial deficit-length framework, capable of decomposing density, diffusivity, and conductivity of confined electrolyte solutions into interfacial and confinement contributions. By applying the framework to molecular dynamics simulations of aqueous alkali halides in planar graphene nanoslits coupled to bulk reservoirs at variable electrolyte concentration, we show that genuine confinement effects emerge only for slit heights \(H\lesssim\SI{1}{nm}\); for larger \(H\), deviations from bulk behaviour are quantitatively captured by interfacial deficit lengths. These deficit lengths are strongly ion-specific and generally positive for water and salt densities as well as for conductivities, meaning that graphene interfaces reduce the values of these observables, while diffusivity deficit lengths tend to be negative, corresponding to larger slit self-diffusivities relative to the corresponding bulk reference. Our deficit-length framework is applicable to any observable from experiments or simulations on nanoconfined electrolytes and predicts confined electrolyte properties for variable slit height H and reservoir electrolyte concentrations.
    \end{abstract}

  \vspace{0.5em}
  \vspace{1em}
]

\begingroup
\renewcommand\thefootnote{*}
\footnotetext{netz@physik.fu-berlin.de}
\endgroup

\FloatBarrier
Electrolytes in nanometre confinement exhibit altered compositions~\cite{luo2015,fang2019,liu2023}, transport properties~\cite{cheng2016,fong2025}, and thermodynamic and dielectric responses~\cite{Schlaich2016PlanarConfinement, gao2018,fumagalli2018} relative to their bulk counterparts. This confinement dependence can be harnessed in applications ranging from energy storage~\cite{chmiola2006,simon2008} and desalination~\cite{cohenTanugi2012} to voltage-controlled nanofluidic transport~\cite{mouterde2019}. There are multiple origins for the observed deviations from bulk behaviour. The amount of water entering a slit varies with slit height and surface properties~\cite{calero2020,wei2022}. Once inside, the water forms pronounced interfacial layers~\cite{iiyama1995,argyris2008,garcia2023,Sam2026Polarization} and exhibits a strongly anisotropic dielectric response near the interface that approaches bulk-like behaviour further away~
\cite{Schlaich2016PlanarConfinement,fumagalli2018,
Loche2020ElectrostaticsNanoconfinement}. When nanoslits are reduced to molecular dimensions, ion entry and selectivity can be governed by steric constraints~
\cite{esfandiar2017}, partial dehydration~
\cite{Yu2019Dehydration} and ion-specific interactions with layered confined water~\cite{goutham2023} or regulated by surface charge and Donnan partitioning~\cite{kalluri2011,Kim2024Donnan}, producing slit electrolyte compositions that differ from those of the connected bulk reservoirs and, in extreme cases, complete ion exclusion~\cite{gopinadhan2019}.

Because many of the observed modifications are tied to the presence of interfacial regions in nanoslits~\cite{bocquet2010,zhan2019,Ntim2020ImageCharges}, it remains difficult to distinguish confinement from interfacial effects~\cite{kavokine2021}. In fact, changes of properties in confinement may arise solely from the varying proportion of interfacial and bulk-like regions within the slit~\cite{Becker2024InterfacialConfinement,Ntim2024Capacitance,Wang2025Interfaces}. This is particularly evident in transport studies, where confinement has been reported to modify diffusivities~\cite{lai2004,Pean2015Confinement,kong2017,shao2021,rezlerova2023} and ionic conductivity~\cite{zhou2021,Emmerich2022Nanofluidic,fong2025}, depending on the channel specificities, electrolyte type, and observable. 

To separate confinement from interfacial contributions, we introduce a deficit-length framework based on the Gibbs dividing-surface concept~\cite{Gibbs1878Heterogeneous} and apply it to water and salt concentrations, self-diffusion, and ionic conductivity of graphene slit--reservoir systems from molecular dynamics (MD) simulations. We use graphene as a model system as it provides chemically simple, atomically flat surfaces and permits slit channels with heights controlled down to the {\aa}ngstr\"om scale~\cite{geim2021,radha2016}. 

\begin{figure*}[!t] \centering \includegraphics[width=\textwidth]{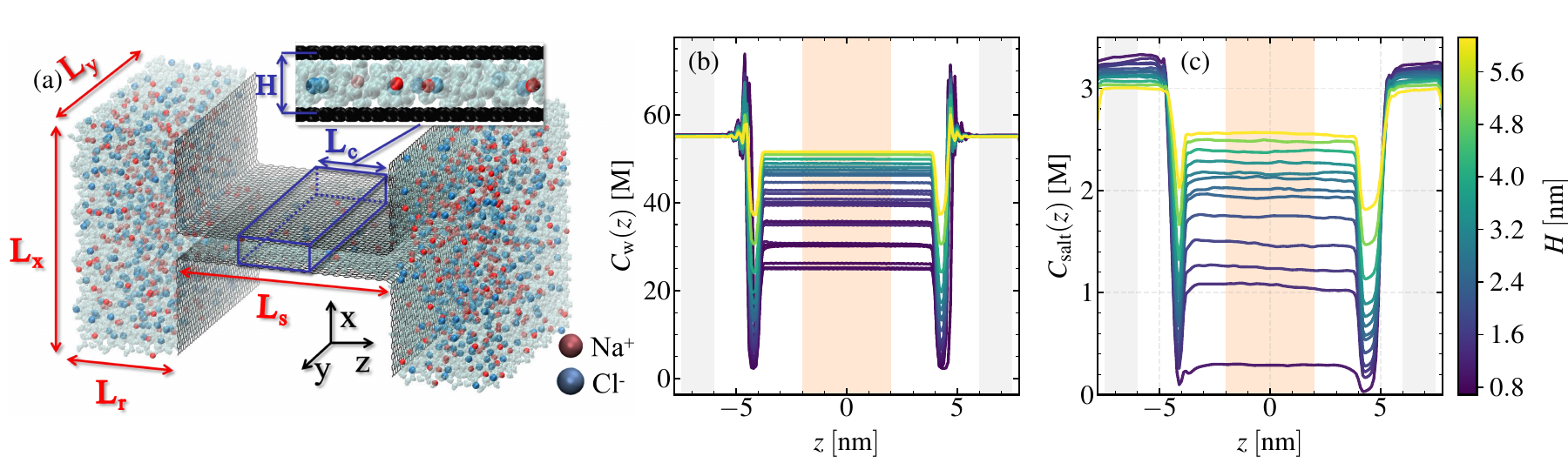} 
\caption[Slit--reservoir simulation geometry]{Slit--reservoir simulation geometry. \textbf{(a)} Snapshot for slit height \(H=\SI{1.00}{nm}\) and average NaCl concentration \(C_{\rm salt}^{\rm ave}\approx\SI{3}{M}\). 
Two reservoirs (\(L_x\times L_y\times L_r\)) are connected by a planar graphene slit of length \(L_s\), depth \(L_y\), and height \(H\). \textbf{(b)} Water molar concentration profile \(C_{\rm w}(z)\) along \(z\) for pure-water systems for different \(H\). \textbf{(c)} Salt molar concentration profile \(C_{\rm salt}(z)\) along \(z\) for NaCl systems with different \(H\) and \(C_{\rm salt}^{\rm ave}\approx\SI{3}{M}\). In (b,c), the central shaded orange area indicates the window used to calculate \(C_{\rm w}^{\rm slit}\) and \(C_{\rm salt}^{\rm slit}\), while the two outer shaded grey areas indicate the windows used to calculate \(C_{\rm w}^{\rm bulk}\) and \(C_{\rm salt}^{\rm bulk}\).}
\label{fig:system} 
\end{figure*}

We simulate pure water and aqueous 1:1 electrolytes (NaCl, NaI, LiCl) using molecular dynamics (MD) in the slit--reservoir geometry of Fig.~\ref{fig:system}a, where two bulk reservoirs are connected by a planar frozen graphene slit of height \(H\), defined as the carbon-plane centre-to-centre distance. From the profiles in Fig.~\ref{fig:system}b,c, we obtain the slit and bulk molar concentrations \(C_X^{\rm slit}\) and \(C_X^{\rm bulk}\), with \(X=w, salt\) for water and ion pairs. Water enters the central slit for \(H\ge\SI{0.65}{nm}\) and salt does so for \(H\ge\SI{1.0}{nm}\). For the electrolyte simulations, we prepare systems with system-wide average salt concentrations \(C_{\rm salt}^{\rm ave}\) of approximately \(0.2,\,1,\,2,\,3,\,4,\) and \(\SI{5}{M}\). These system-wide averages differ slightly from the reservoir concentrations \(C_{\rm salt}^{\rm bulk}\), measured after equilibration in the NPT ensemble. Further simulation details are given in SI~Secs.~\ref{si:system} and~\ref{si:md}.

The central idea of our analysis is to express each observable in the slit \(\Theta^{\rm slit}\) in terms of the bulk reference value
\(\Theta^{\rm bulk}\) according to
\begin{equation}
  \Theta^{\rm slit}/\Theta^{\rm bulk}
  =
  1-2\delta_\Theta(H)/H
  \approx
  1-2\delta_\Theta^{\infty}/H ,
  \label{eq:ratio}
\end{equation}
where \(\Theta\in\{C_w,C_{\rm salt},D_-,D_+,D_w,\sigma\}\) is a placeholder for the water and salt concentrations \((C_w,C_{\rm salt})\), the anion, cation, and water self-diffusion coefficients \((D_-,D_+,D_w)\), and the ionic conductivity \((\sigma)\); \(\delta_\Theta(H)\) denotes the \(H\)-dependent deficit length and constitutes a generalization of the Gibbs deficit length introduced for the density deficit at a single interface~\cite{Gibbs1878Heterogeneous,Mamatkulov2004Hydrophobic,janecek2007interfacial} to diffusional and transport properties in confinement. We define intensive slit quantities \(\Theta^{\rm slit}\) by dividing the corresponding extensive quantities by the slit height \(H\), exactly as one would proceed in an experimental measurement. Our main finding is that for \(H \gtrsim \SI{1}{nm}\) \(\delta_\Theta(H) \approx \delta_\Theta^\infty\) holds very well for all observables studied and thus interfacial effects dominate over confinement effects.

\begin{figure*}[!t]
  \centering
  \includegraphics[width=\textwidth]{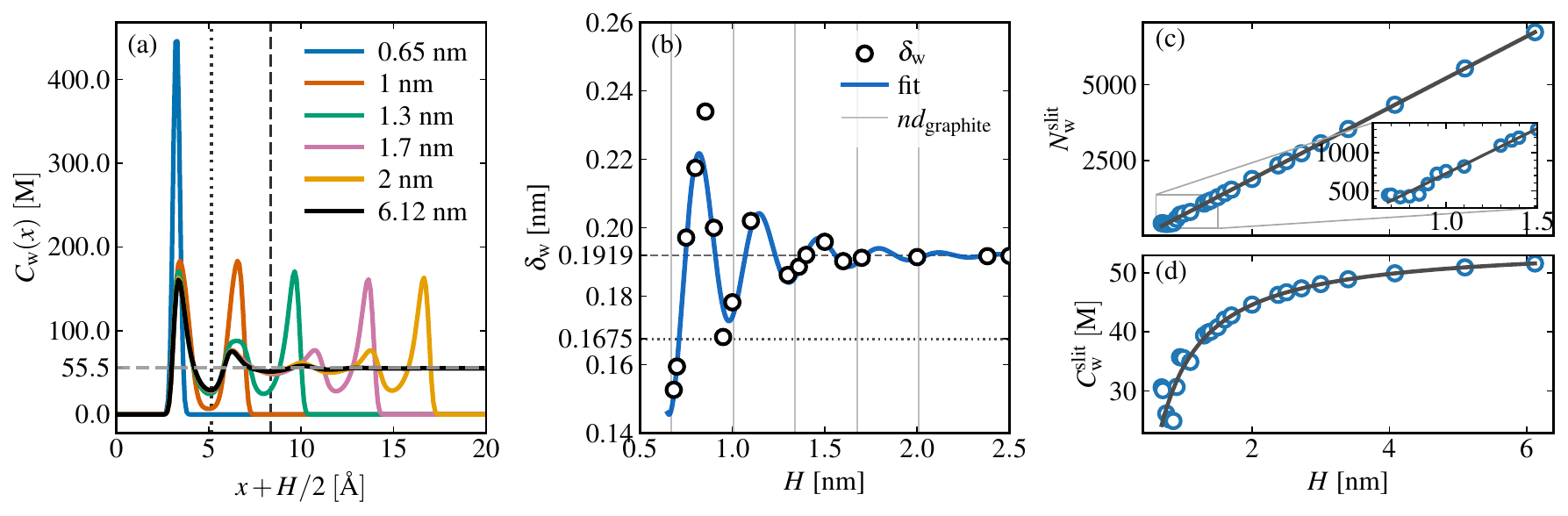}
\caption[Results for confined pure water]{Results for confined pure water.
  \textbf{(a)} Water density profiles \(C_w(x)\) for representative \(H\) values, plotted against \(x+H/2\), so that \(x+H/2=0\) corresponds to the position of a graphene wall. Vertical lines mark the first two minima of the  \(H=\SI{6.12}{nm}\) system, \(\ell_1=\SI{0.515}{nm}\) and \(\ell_2=\SI{0.835}{nm}\); the horizontal dashed line is the bulk water molarity  \(C_w^{\rm bulk}\) = \(\SI{55.5}{M}\).
  \textbf{(b)} Deficit length \(\delta_w(H)\) defined by Eq.~\eqref{eq:ratio}, with a damped-cosine fit
  [Eq.~\eqref{eq:damped_cosine}; SI~Sec.~\ref{si:dwfit}]. Dashed and dotted horizontal lines indicate
  \(\delta_w^{\infty}=\SI{0.1919}{nm}\) and half the graphite interlayer spacing \(d_{\rm graphite}/2=0.335/2=\SI{0.1675}{nm}\)~\cite{Franklin1951GraphiticCarbons},
  respectively; grey vertical lines mark multiples of the graphite interlayer spacing \(n\,d_{\rm graphite}\).
  \textbf{(c)} Mean slit water count \(N_w^{\rm slit}(H)\) compared with the asymptotic deficit-length prediction \(C_w^{\rm bulk}N_A L_yL_c(H-2\delta_w^\infty)\) (see SI~Sec.~\ref{si:system} for details).
  \textbf{(d)} Slit water molarity \(C_w^{\rm slit}(H)\) compared with the asymptotic deficit-length prediction \(C_w^{\rm bulk}(1-2\delta_w^\infty/H)\).}
  \label{fig:pure_water}
\end{figure*}

We start by establishing the density deficit length for pure water in graphene nanoslits. The profiles \(C_w(x)\) in Fig.~\ref{fig:pure_water}a show pronounced interfacial oscillations that reflect water layering and settle to the bulk molarity \(\SI{55.5}{M}\). In the widest slit, \(H=\SI{6.12}{nm}\), the first two minima occur at \(\ell_1=\SI{0.515}{nm}\) and \(\ell_2=\SI{0.835}{nm}\) from the wall. The first interfacial water layers on the opposing walls meet for \(H\approx2\ell_1=\SI{1.03}{nm}\) and give rise to a perfectly structured two-layer state, as seen for \(H=\SI{1.00}{nm}\) in Fig.~\ref{fig:pure_water}a.

The deficit length \(\delta_w(H)\) extracted from Eq.~\eqref{eq:ratio} in Fig.~\ref{fig:pure_water}b oscillates and reaches a plateau for \(H>\SI{2}{nm}\). Its minima reflect slit heights that fit a whole number of water layers---efficient packing, hence little apparent depletion---whereas its maxima mark frustrated, partially filled slits. The sampled heights \(H=0.65,\,1.00,\,1.30,\) and \(\SI{1.70}{nm}\) therefore approximately correspond to one-, two-, three-, and four-layer water structures. The deficit length is well fit by
\begin{equation}
  \delta_w(H)
  \approx
  \delta_w^{\infty}
   +A_0\,\exp(-H/\xi)\,
        \cos\!\bigl[2\pi(H-H_0)/\lambda\bigr],
  \label{eq:damped_cosine}
\end{equation}
with period \(\lambda=\SI{0.325}{nm}\), which matches well the graphite
layer spacing \(d_{\rm graphite}=\SI{0.335}{nm}\)~\cite{Franklin1951GraphiticCarbons}
(see SI~Sec.~\ref{si:dwfit}).

When the two interfacial regions are sufficiently separated, increasing \(H\) mainly adds bulk-like interior, while each interface contributes approximately the same deficit and \(\delta_w(H)\) approaches the large-\(H\) deficit length \(\delta_w^{\infty}=\SI{0.1919}{nm}\) where  the large-\(H\) approximation in Eq.~\eqref{eq:ratio} becomes applicable, in agreement with previous results~\cite{Loche2020ElectrostaticsNanoconfinement}. Incidentally, \(2\delta_w^\infty=\SI{0.384}{nm}\) is close to \(d_{\rm graphite}\) and to \(\lambda\), which explains why a single layer of water enters readily into a slit produced by removing one graphene sheet from graphite~\cite{Smith2024CapillaryWettability,gopinadhan2019,radha2016}.

Using the obtained \(\delta_w^{\infty}\), the approximate concentration prediction from Eq.~\eqref{eq:ratio},  \(C_w^{\rm slit}(H)\approx C_w^{\rm bulk}(1-2\delta_w^{\infty}/H)\), describes both the mean slit water number and the corresponding slit molarity very well (Fig.~\ref{fig:pure_water}c,d), with only small deviations in the strongly layered sub-nanometre regime, which demonstrates the usefulness of the deficit-length framework. The inset of Fig.~\ref{fig:pure_water}c shows that the deviations correspond to discrete layer filling in narrow slits: single-layer water below \(H\approx\SI{0.85}{nm}\) and two water layers near \(H=\SI{1.00}{nm}\).

\begin{figure*}[!t]
  \centering
  \includegraphics[width=\linewidth]{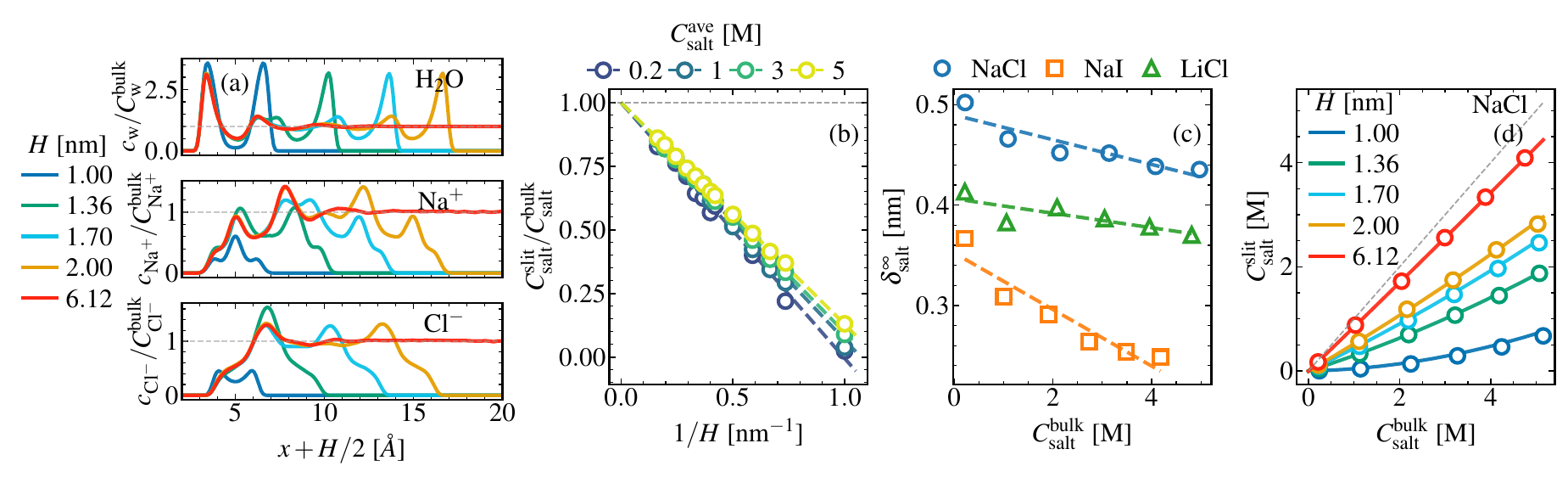}
  \caption[Results for confined electrolytes]{Results for confined electrolytes.
  \textbf{(a)} Normalised density profiles for NaCl at \(C_{\rm salt}^{\rm ave}\approx\SI{5}{M}\) for representative slit heights \(H\): from top to bottom, \(C_w(x)/C_w^{\rm bulk}\), \(C_{\mathrm{Na}^+}(x)/C_{\mathrm{Na}^+}^{\rm bulk}\), and \(C_{\mathrm{Cl}^-}(x)/C_{\mathrm{Cl}^-}^{\rm bulk}\), plotted against \(x+H/2\).
  \textbf{(b)} Slit-to-bulk salt concentration ratio
  \(C_{\rm salt}^{\rm slit}/C_{\rm salt}^{\rm bulk}\) for NaCl as a
  function of \(1/H\) at different \(C_{\rm salt}^{\rm ave}\). Dashed lines are fixed-intercept fits of Eq.~\eqref{eq:ratio} for \(H\ge\SI{2.00}{nm}\), used to extract \(\delta_{\rm salt}^{\infty}\).
  \textbf{(c)} Concentration-dependent salt deficit length \(\delta_{\rm salt}^{\infty}(C_{\rm salt}^{\rm bulk})\) for NaCl, NaI, and LiCl. For each series, \(C_{\rm salt}^{\rm bulk}\) is the mean reservoir concentration over the sampled slit heights. Dashed lines are linear fits to the extracted \(\delta_{\rm salt}^{\infty}\) values and provide the concentration-dependent input used in panel (d) for NaCl.
  \textbf{(d)} Slit salt molarity \(C_{\rm salt}^{\rm slit}\) predicted
  from Eq.~\eqref{eq:ratio} using the linear fits to
  \(\delta_{\rm salt}^{\infty}(C_{\rm salt}^{\rm bulk})\) from panel
  (c), compared with simulation data (symbols) for different slit
  heights.}
  \label{fig:nacl_summary}
\end{figure*}

We next extend the deficit-length analysis to electrolyte solutions. The NaCl density profiles at \(C_{\rm salt}^{\rm ave}\approx\SI{5}{M}\) for representative slit heights \(H\) in Fig.~\ref{fig:nacl_summary}a show that at \(H=\SI{1.00}{nm}\), \ce{Na+} and \ce{Cl-} are restricted to the narrow region between the two water layers. As \(H\) increases, the accessible region broadens; by \(H\ge\SI{2.00}{nm}\), a bulk-like region separates the two interfacial domains. We therefore fit Eq.~\eqref{eq:ratio} to \(C_{\rm salt}^{\rm slit}/C_{\rm salt}^{\rm bulk}\) for \(H\ge\SI{2.00}{nm}\), to obtain \(\delta_{\rm salt}^{\infty}\) for each \(C_{\rm salt}^{\rm ave}\) as shown for NaCl in Fig.~\ref{fig:nacl_summary}b. At fixed \(C_{\rm salt}^{\rm ave}\), the reservoir concentration \(C_{\rm salt}^{\rm bulk}\) varies slightly with \(H\), as shown for NaCl in SI~Fig.~\ref{fig:water_electrolyte}b. Since each \(\delta_{\rm salt}^{\infty}\) is extracted from an \(H\)-series, in
Fig.~\ref{fig:nacl_summary}c we plot \(\delta_{\rm salt}^{\infty}\) over the mean of
\(C_{\rm salt}^{\rm bulk}\) for each fixed
\(C_{\rm salt}^{\rm ave}\). For NaCl, NaI, and LiCl, the resulting deficit lengths decrease approximately linearly with increasing reservoir concentration (Fig.~\ref{fig:nacl_summary}c, dashed lines) and are strongly salt-specific. The common decrease indicates that, for the three ion pairs studied here, interfacial ion depletion becomes less pronounced as \(C_{\rm salt}^{\rm bulk}\) increases. At lower concentrations, the linear trend observed is expected to change into a scaling proportional to the electrolyte screening length, i.e. inversely proportional to the square root of the salt concentration~\cite{dosSantos2020IonSpecificity}.

The approximation of Eq.~\eqref{eq:ratio}  \(C_{\rm salt}^{\rm slit}\approx C_{\rm salt}^{\rm bulk}(1-2\delta_{\rm salt}^{\infty}/H)\),
using \(\delta_{\rm salt}^{\infty}(C_{\rm salt}^{\rm bulk})\) from the linear fits in Fig.~\ref{fig:nacl_summary}c, reproduces the slit salt concentration for all bulk salt concentrations and slit heights (Fig.~\ref{fig:nacl_summary}d): narrow slits hold less salt, wide slits approach the reservoir concentration (dashed line). The corresponding results for water are shown in SI~Fig.~\ref{fig:water_electrolyte}. We conclude that for \(H>\SI{1}{nm}\), i.e., beyond the two-water-layer regime, the slit water and salt concentrations are very well described by the ion-specific, concentration-dependent deficit lengths \(\delta_w^{\infty}\) and \(\delta_{\rm salt}^{\infty}\).

\begin{figure*}[!t]
  \centering
  \includegraphics[width=\linewidth]{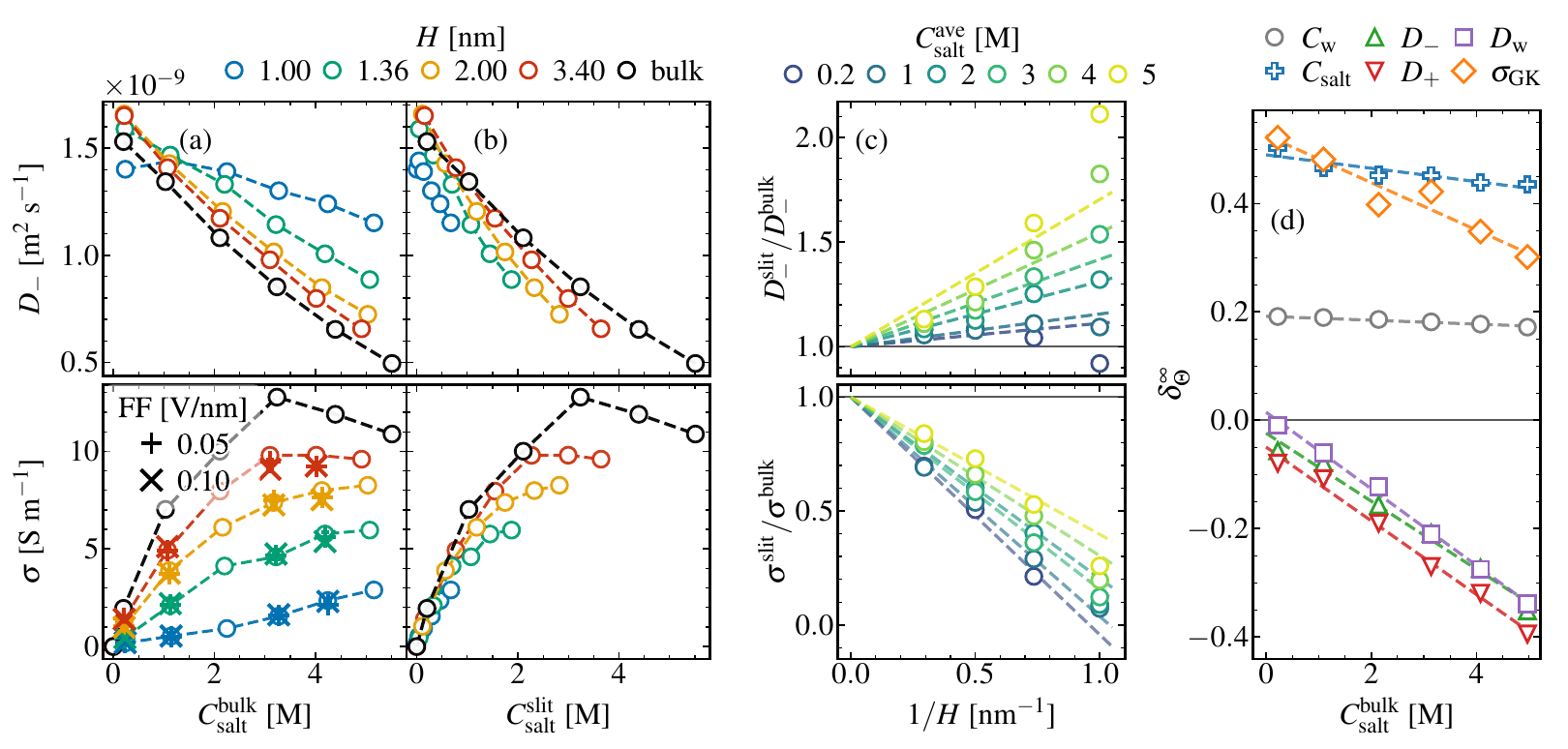}
\caption[NaCl transport properties in slit-only simulations]{NaCl transport properties in slit-only simulations. 
  \textbf{(a)} Anion self-diffusion \(D_-\) (top) and ionic conductivity \(\sigma\) (bottom) in the slit for different \(H\), plotted against their reservoir salt molarity \(C_{\rm salt}^{\rm bulk}\). Coloured open circles show slit simulation data; black open circles denote the bulk reference simulation results. For conductivity, the open circles correspond to \(\sigma_{\rm GK}\), obtained using the Green--Kubo relation, while \(\sigma_{\rm FF}^{0.05}\) and \(\sigma_{\rm FF}^{0.10}\) are finite-field estimates at \(E=\SI{0.05}{V/nm}\) and \(\SI{0.10}{V/nm}\) (plus signs and crosses, respectively).
  \textbf{(b)} Same transport observables plotted against the slit salt molarity \(C_{\rm salt}^{\rm slit}\), with the bulk-reference data plotted as a function of \(C_{\rm salt}^{\rm bulk}\) for comparison.
  \textbf{(c)} Slit-to-bulk ratios versus \(1/H\) for \(D_-\) (top) and \(\sigma\) (bottom). The transport bulk-reference values are obtained by interpolation from separate bulk simulations (SI~Sec.~\ref{si:slitonly}). 
  \textbf{(d)} Large-\(H\) deficit lengths \(\delta_\Theta^{\infty}(C_{\rm salt}^{\rm bulk})\) for \(\Theta\in\{C_w,C_{\rm salt},D_-,D_+,D_w,\sigma\}\). For each value of \(C_{\rm salt}^{\rm ave}\), the mean \(C_{\rm salt}^{\rm bulk}\) is obtained by averaging the measured reservoir concentrations over the sampled \(H\) shown in SI~Fig.~\ref{fig:water_electrolyte}b. Dashed lines are linear fits of Eq.~\eqref{eq:deltaC}, numerical coefficients are listed in SI~Table~\ref{tab:delta_coeffs}.}
  \label{fig:transport_nacl}
\end{figure*}

Now we turn to in-plane electrolyte transport properties in graphene slits obtained from slit-only simulations using the water and salt concentrations obtained from our slit-reservoir simulations. Fig.~\ref{fig:transport_nacl}a shows the \(\ce{Cl-}\) self-diffusivity \(D_-\) (top) and the conductivity \(\sigma\) (bottom) as functions of \(C_{\rm salt}^{\rm bulk}\) for NaCl, where we observe for both properties a pronounced dependence on the slit height \(H\). The conductivity is evaluated from the Green--Kubo relation (open circles), with a few reference points calculated also using finite-field non-equilibrium molecular dynamics (NEMD) (plus signs and crosses). Their agreement (Fig.~\ref{fig:transport_nacl}a lower panel) indicates that the applied fields, \(E=\SI{0.05}{V/nm}\) and \(\SI{0.10}{V/nm}\), are in the linear-response regime. Simulation and method details are given in SI~Secs.~\ref{si:slitonly}--\ref{si:ff}. Replotting the same data against \(C_{\rm salt}^{\rm slit}\) in Fig.~\ref{fig:transport_nacl}b brings both \(D_-\) and \(\sigma\) values closer to their bulk values. This suggests that a substantial part of the \(H\)-dependence in Fig.~\ref{fig:transport_nacl}a arises from concentration differences among the slits, though it is important to realise that the partial data collapse in Fig.~\ref{fig:transport_nacl}b depends on the definition of the slit variables \(D_-\) and \(\sigma\), which we arbitrarily base on the slit height \(H\) defined by the graphene carbon-atom separation. Results for cation and water diffusivities \(D_+\) and \(D_w\) are shown in SI~Fig.~\ref{fig:si_selfD_cbulk_cslit}.

We extract the transport deficit lengths \(\delta_{D_-}^{\infty}\) and
\(\delta_{\sigma}^{\infty}\) from linear fits of the slit-to-bulk ratios \(D_-^{\rm slit}/D_-^{\rm bulk}\) and \(\sigma^{\rm slit}/\sigma^{\rm bulk}\) versus \(1/H\) in Fig.~\ref{fig:transport_nacl}c for \(H>\SI{1}{nm}\) and for each \(C_{\rm salt}^{\rm ave}\). The bulk-reference values \(D_-^{\rm bulk}\) and
\(\sigma^{\rm bulk}\) are evaluated at the reservoir concentration \(C_{\rm salt}^{\rm bulk}\) by interpolation, details are given in SI~Sec.~\ref{si:slitonly}. In Fig.~\ref{fig:transport_nacl}d we compare the deficit lengths as functions of \(C_{\rm salt}^{\rm bulk}\). The concentration and conductivity deficit lengths (\(\delta_w^{\infty}\), \(\delta_{\rm salt}^{\infty}\), \(\delta_{\sigma}^{\infty}\)) are positive, reflecting genuine deficits, and decrease slightly and  approximately linearly with \(C_{\rm salt}^{\rm bulk}\), whereas the self-diffusion deficit lengths (\(\delta_{D_-}^{\infty}\), \(\delta_{D_+}^{\infty}\), \(\delta_{D_w}^{\infty}\)) are negative over the sampled concentration range and decrease more strongly. The close agreement between \(\delta_{\rm salt}^{\infty}\) and \(\delta_{\sigma}^{\infty}\) at low to moderate concentrations suggests that changes in the number of charge carriers entering the slit make an important contribution to the conductivity deficit. The negative self-diffusion deficit lengths reflect an enhanced
diffusivity at the graphene-water interface due to
slip~\cite{bocquet2010,carlson2025_interfacial}.

\begin{figure*}[!t]
  \centering
  \includegraphics[width=\linewidth]{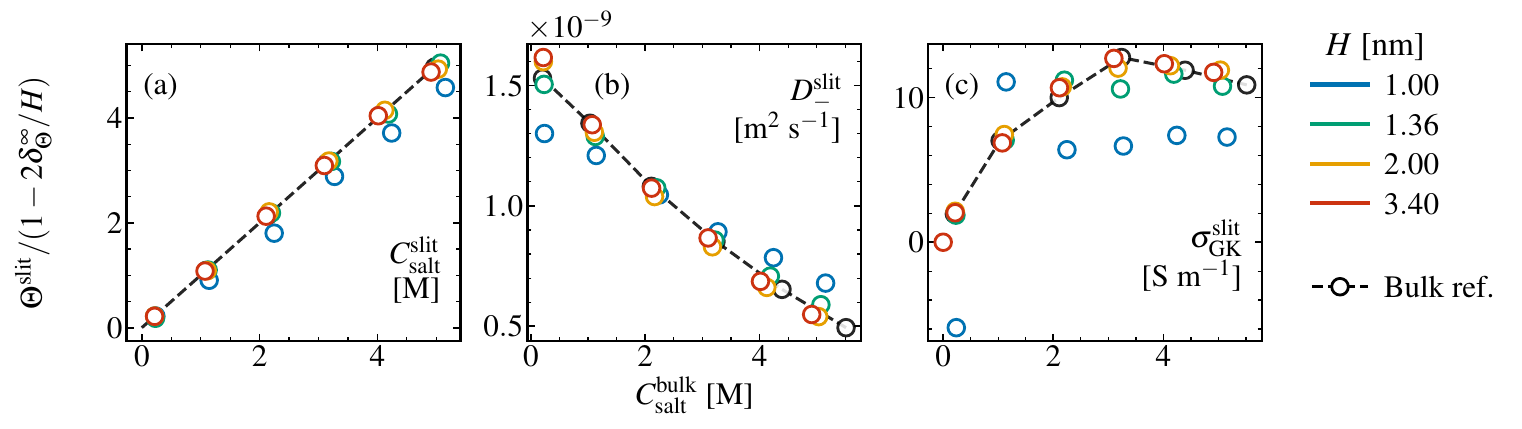}
    \caption[Universal plot of rescaled slit observables]{Universal plot of rescaled slit observables
    \(\Theta^{\rm slit}/(1-2\delta_\Theta^{\infty}/H)\) for confined aqueous NaCl.
    \textbf{(a)} Slit NaCl concentration \(C_{\rm salt}^{\rm slit}\).
    \textbf{(b)} Anion self-diffusion \(D_-^{\rm slit}\).
    \textbf{(c)} Ionic conductivity \(\sigma^{\rm slit}\). Coloured circles show the slit data, compared with bulk reference data (black circles connected by black broken lines). Colours indicate slit height \(H\). Deviations from the bulk reference data are only present for \(H = 1.0 \) nm. Other observables, \(C_w^{\rm slit}\), \(D_+^{\rm slit}\), and \(D_w^{\rm slit}\), are shown in SI~Fig.~\ref{fig:transport_corrected_si}.}
  \label{fig:transport_corrected}
\end{figure*}

Having expressed composition and transport in confinement by deficit lengths, we test whether a universal description of electrolyte properties in slits is possible. For this, we describe the concentration dependence of the deficit lengths by the linear expression
\begin{equation}
  \delta_\Theta^{\infty}\!\bigl(C_{\rm salt}^{\rm bulk}\bigr)
  =
  \alpha_\Theta+\beta_\Theta\,C_{\rm salt}^{\rm bulk},
  \label{eq:deltaC}
\end{equation}
where for the water-concentration we fix \(\delta_w^{\infty}(C_{\rm salt}^{\rm bulk}=0)
=\SI{0.1919}{nm}\), obtained from Fig.~\ref{fig:pure_water}b; all other coefficients follow from fits as shown in Fig.~\ref{fig:transport_nacl}d.

In Fig.~\ref{fig:transport_corrected}, we use the fitted \(\delta_\Theta^{\infty}(C_{\rm salt}^{\rm bulk})\) from Eq.~\eqref{eq:deltaC} to rescale each slit observable as \(\Theta^{\rm slit}/(1-2\delta_\Theta^{\infty}/H)\) according to Eq.~\eqref{eq:ratio} and compare \(C_{\rm salt}^{\rm slit}\), \(D_-^{\rm slit}\), and \(\sigma^{\rm slit}\) with their bulk reference data (black broken lines, other observables are shown in SI~Fig.~\ref{fig:transport_corrected_si}). For all observables, the rescaled data lie perfectly on the bulk reference for \(H\ge\SI{1.36}{nm}\), while systematic deviations are seen for \(H=\SI{1.00}{nm}\), meaning that confinement effects are only significant for \(H\lesssim\SI{1}{nm}\). Thus, the deficit-length framework yields a universal definition and description of interfacial and confinement effects on electrolyte densities and density-dependent transport properties.

In summary, we apply a Gibbs-like interfacial deficit-length description to molecular-dynamics simulations of pure water and aqueous 1:1 alkali halides in graphene slit--reservoir systems to separate interfacial and confinement contributions to concentrations and in-plane transport properties. The deficit lengths are positive for concentrations and the conductivity, but negative and strongly concentration-dependent for the self-diffusivity. After deficit-length rescaling, all observables collapse onto their bulk references for \(H\ge\SI{1.36}{nm}\), whereas small deviations are present at \(H=\SI{1.00}{nm}\). This indicates that once the slit accommodates more than two water layers, the effect of confinement is largely captured by the interfacial contributions. The onset of genuine confinement effects for \(H \lesssim \SI{1.0}{nm}\) is consistent with previous structural and dielectric-response studies~\cite{Schlaich2016PlanarConfinement, Fellows2024AirWaterThickness, Becker2024InterfacialConfinement, Wang2025Interfaces, Lehmann2026MultipolarSFG}. Overall, the deficit-length framework suggests that much of the slit-height dependence commonly attributed to confinement reflects the changing balance between interfacial and bulk-like solution, while genuine confinement emerges when the two interfacial regions overlap and no bulk-like region remains, that is, for \(H\lesssim\SI{1}{nm}\). In fact, the deficit length can be directly extracted from measurements of the property under study as a function of slit height, our framework can thus be applied to experimental data as well as simulation data from classical force-field, ab initio or machine-learning potentials that may also include quantum nuclear effects.

\FloatBarrier
\clearpage
\FloatBarrier
\clearpage
\bibliographystyle{unsrtnat}
\bibliography{ion_confinement}


\onecolumn

\setcounter{section}{0}
\setcounter{figure}{0}
\setcounter{table}{0}
\setcounter{equation}{0}

\renewcommand{\thesection}{S\arabic{section}}
\renewcommand{\thefigure}{S\arabic{figure}}
\renewcommand{\thetable}{S\arabic{table}}
\renewcommand{\theequation}{S\arabic{equation}}

\renewcommand{\theHsection}{S\arabic{section}}
\renewcommand{\theHfigure}{S\arabic{figure}}
\renewcommand{\theHtable}{S\arabic{table}}
\renewcommand{\theHequation}{S\arabic{equation}}


\begin{center}

{\LARGE\bfseries Supporting Information\par}

\vspace{1.3em}

{\Large\bfseries
Electrolyte density, diffusivity and conductivity in graphene nanoconfinement predicted by separating interfacial from genuine confinement effects
\par}

\vspace{1.2em}

{\normalsize
Haoyuan Quan\textsuperscript{1},
Hanne S. Antila\textsuperscript{1},
Maximilian R. Becker\textsuperscript{1},
Philip R. Loche\textsuperscript{2},
and Roland R. Netz\textsuperscript{1,*}
\par}

\vspace{0.7em}

{\normalsize\itshape
\textsuperscript{1}Freie Universit\"at Berlin, Department of Physics,
Institute for Theoretical Physics, Arnimallee 14,
14195 Berlin, Germany
\par}

\vspace{0.25em}

{\normalsize\itshape
\textsuperscript{2}Laboratory of Computational Science and Modeling,
Institut des Mat\'eriaux, \'Ecole Polytechnique F\'ed\'erale de Lausanne,
1015 Lausanne, Switzerland
\par}

\vspace{0.6em}

{\normalsize
\textsuperscript{*}\texttt{netz@physik.fu-berlin.de}
\par}

\end{center}

\vspace{1.5em}

\section{Simulation box}
\label{si:system}

The simulation box contains two reservoirs connected by a planar graphene slit. The slit normal is \(x\), and the graphene sheets lie in the \(y\)--\(z\) plane. The graphene slit height \(H\) is defined as the carbon-plane centre-to-centre distance. The initial reservoir dimensions are \(L_x\times L_y\times L_r=\SI{8.52}{nm}\times\SI{8.85}{nm} \times\SI{4.00}{nm}\), where \(L_r\) is the initial reservoir length along \(z\); the slit has length \(L_s=\SI{8.52}{nm}\) along \(z\) and width \(L_y=\SI{8.85}{nm}\) along \(y\). The central slit window length is \(L_c=\SI{4.00}{nm}\) along \(z\), as shown by the orange shaded region in Fig.~\ref{fig:system}b,c, and the two bulk plateau windows have length \(L_{\rm bulk}=\SI{1.00}{nm}\) each, as shown by the grey shaded regions. All slit and bulk concentrations reported in the main text are averaged over these central-slit and bulk plateau windows over the final
\(\SI{200}{ns}\).

For pure water, the water deficit-length analysis includes slit heights \(H=\SIrange{0.65}{6.12}{nm}\). For \(H<\SI{0.60}{nm}\), the central slit is water-free over the analysed trajectory and is therefore not included in the water deficit-length analysis. For NaCl, the salt deficit-length analysis includes \(H=\SIrange{1.00}{6.12}{nm}\), and for LiCl and NaI it includes \(H=\SIrange{1.00}{4.08}{nm}\). For \(H<\SI{1.00}{nm}\), the central slit window is ion-free and is therefore not included in the salt deficit-length analysis.

Molar concentrations in the slit and bulk analysis windows are computed
from the number of particles in the corresponding analysis volume,
\begin{equation}
\begin{aligned}
  C_X^{\rm slit}
  &=
  \frac{N_X^{\rm slit}}{N_A L_yL_cH},
  \\
  C_X^{\rm bulk}
  &=
  \frac{N_X^{\rm bulk}}{N_A(2L_{\rm bulk}L_xL_y)} .
\end{aligned}
\label{eq:window_concentrations}
\end{equation}
Here, \(X\in\{\mathrm{w},\mathrm{salt}\}\), \(N_A\) is Avogadro's constant, and molar concentrations are reported in \(\mathrm{M}=\mathrm{mol\,L^{-1}}\). Thus, \(C_{\rm salt}^{\rm bulk}\) is the salt concentration
in the reservoir plateau windows rather than the system-wide average salt concentration used to prepare the simulation. At fixed
\(C_{\rm salt}^{\rm ave}\), \(C_{\rm salt}^{\rm bulk}\) therefore varies slightly with slit height \(H\), as shown for NaCl in Fig.~\ref{fig:water_electrolyte}b.

The profiles along z in Fig.~\ref{fig:system}b,c are computed
by binning particles along \(z\) with slab thickness
\(\Delta z=\SI{1.0}{\angstrom}\). The local molar concentration
\(C_X(z)\) is computed from the particle count in each slab divided by
the corresponding local slab volume.

For 1:1 salts, the salt count in each analysis window is computed as
the mean of the cation and anion counts,
\(
  N_{\rm salt}
  =
  \frac{1}{2}
  \left(N_{\rm cation}+N_{\rm anion}\right).
\)
For the local salt profile \(C_{\rm salt}(z)\), only locally neutral
cation--anion pairs are counted in each slab, according to
\(
  N_{\rm salt}(z;\Delta z)
  =
  \min\!\left[
  N_{\rm cation}(z;\Delta z),
  N_{\rm anion}(z;\Delta z)
  \right].
\)

The system-wide average salt concentration is reported as
\(
  C_{\rm salt}^{\rm ave}
  ={n_{\rm salt}}/{n_{\rm w}}\times\SI{55.5}{M},
\)
where \(n_{\rm salt}\) and \(n_{\rm w}\) are the numbers of ion pairs
and water molecules initially placed in the entire simulation system,
respectively, and \(\SI{55.5}{M}\) is the molar concentration of pure
water used as the reference. Example initial compositions for the
\(H=\SI{1.00}{nm}\) NaCl systems are listed in
Table~\ref{tab:composition}.

\begin{table}[h]
  \centering
  \caption[Example initial compositions for NaCl systems]{Example initial compositions for \(H=\SI{1.00}{nm}\) NaCl
  systems at different system-wide average salt concentrations.}
  \label{tab:composition}
  \begin{tabular}{rrr}
    \toprule
    \(n_{\rm salt}\) & \(n_{\rm w}\) & \(C_{\rm salt}^{\rm ave}\) (M) \\
    \midrule
       73 & 20295 & 0.2 \\
      355 & 19731 & 1.0 \\
      687 & 19067 & 2.0 \\
      997 & 18447 & 3.0 \\
     1288 & 17865 & 4.0 \\
     1560 & 17321 & 5.0 \\
    \bottomrule
  \end{tabular}
\end{table}

\section{MD simulation protocol and force field}
\label{si:md}

All molecular dynamics simulations were performed with GROMACS~2025.1
under three-dimensional periodic boundary conditions
\cite{gromacs,gromacs2025_1}, with the graphene sheets frozen in all
directions. Each slit--reservoir system was first relaxed by
steepest-descent energy minimisation for up to \(10^5\) steps.
Production simulations were then propagated with the leap-frog
integrator using a \(\SI{2}{fs}\) time step for up to
\(\SI{1}{\micro\second}\) in the isothermal--isobaric ensemble.

The fluid was maintained at \(\SI{300}{K}\) using the stochastic
velocity-rescaling thermostat with \(\tau_t=\SI{0.5}{ps}\)
\cite{Bussi2007VRescale}; the graphene and fluid were assigned separate
temperature-coupling groups. The \(z\)-direction was coupled to
\(\SI{1}{bar}\) using the stochastic cell-rescaling barostat with
\(\tau_p=\SI{2.0}{ps}\) and
\(\kappa_z=4.5\times10^{-5}\,\si{bar^{-1}}\)
\cite{Bernetti2020CRescale}. The transverse compressibility was set to
zero, so that only the channel-direction box length \(L_z\) fluctuated
while the slit geometry was preserved.

The geometry of the SPC/E water molecules was kept rigid using the
SETTLE algorithm. Short-range interactions were evaluated using the
Verlet cutoff scheme. Long-range electrostatics were treated by
particle-mesh Ewald summation with a real-space cutoff of
\(\SI{0.9}{nm}\), fourth-order interpolation, and a Fourier spacing of
\(\SI{0.1}{nm}\). Lennard-Jones interactions were truncated at
\(\SI{0.9}{nm}\) and shifted to zero at the cutoff using the
potential-shift-Verlet modifier, with no long-range dispersion
correction. The main simulation parameters are summarised in
Table~\ref{tab:md_params}.

\begin{table}[h]
  \centering
  \caption[MD parameters for energy minimisation and equilibration]{MD parameters for energy minimisation (EM) and
  slit--reservoir production.}
  \label{tab:md_params}
  \begin{tabular}{lcc}
    \toprule
    Parameter & EM & Production \\
    \midrule
    Integration algorithm & Steepest descent & Leap-frog MD \\
    Time step & ---  & \(\SI{2}{fs}\) \\
    Maximum number of steps & \(10^5\)  & \(5\times10^8\) \\
    Water-constraint algorithm   & SETTLE   & SETTLE\\
    Neighbour list & $r_{\rm list}\!=\!\SI{0.9}{nm}$ & same \\
    Electrostatics & PME, $r_{\rm coul}\!=\!\SI{0.9}{nm}$, spacing
                    $=\!\SI{0.1}{nm}$ & same \\
    VdW cutoff & $r_{\rm vdw}\!=\!\SI{0.9}{nm}$ & same \\
    Thermostat & --- & \makecell{Stochastic velocity rescaling~\cite{Bussi2007VRescale} \\ 
        ($\tau_t\!=\!\SI{0.5}{ps}$, $T\!=\!\SI{300}{K}$)} \\
    Barostat & --- & \makecell{Stochastic cell rescaling, semi-isotropic~\cite{Bernetti2020CRescale} \\
        ($\tau_p\!=\!\SI{2.0}{ps}$, $p_z\!=\!\SI{1}{bar}$)} \\
    \bottomrule
  \end{tabular}
\end{table}

For water we used the SPC/E model~\cite{spce}, and the graphene carbons were described by the Lennard-Jones parameters $\sigma_{\rm C}=\SI{0.3568}{nm}$ and
$\epsilon_{\rm C}=\SI{0.2966}{kJ\,mol^{-1}}$. Together with the SPC/E oxygen parameters ($\sigma_{\rm OW}=\SI{0.3166}{nm}$, $\epsilon_{\rm OW}=\SI{0.6500}{kJ\,mol^{-1}}$), these give the carbon--oxygen cross interaction $\sigma_{\rm C,OW}=\SI{0.3367}{nm}$ and $\epsilon_{\rm C,OW}=\SI{0.4391}{kJ\,mol^{-1}}$, the graphene--water force-field parameters of Carlson \emph{et al.}~\cite{carlson2024_graphene}, which reproduce a graphene--water contact angle of \(80^\circ\), consistent with experimental data. The ionic Lennard-Jones parameters were taken from Due\~nas-Herrera \emph{et al.}~\cite{duenas-herrera2024} (Table~\ref{tab:lj_params}), which follow the
transferable ion-force-field framework of Loche \emph{et
al.}~\cite{loche2021_transferable}. All non-bonded interactions use the 12--6 Lennard-Jones form
\begin{equation}
  U_{ij}(r)=4\epsilon_{ij}\!\left[
   \left(\frac{\sigma_{ij}}{r}\right)^{12}
   -\left(\frac{\sigma_{ij}}{r}\right)^{6}
  \right],
  \qquad r<r_{\rm cut}=\SI{0.9}{nm},
\end{equation}
with the cross terms obtained from generalised Lorentz--Berthelot combining
rules,
\begin{equation}
  \sigma_{ij}=\lambda_\sigma^{ij}\frac{\sigma_i+\sigma_j}{2},
  \qquad
  \epsilon_{ij}=\lambda_\epsilon^{ij}\sqrt{\epsilon_i\epsilon_j},
  \qquad
  \lambda_\sigma^{ij}=\lambda_\epsilon^{ij}=1 .
\end{equation}

\begin{table}[h]
  \centering
  \caption[LJ parameters and ionic charges]{LJ parameters and ionic charges used for the
  alkali and halide ions~\cite{duenas-herrera2024}.}
  \label{tab:lj_params}
  \sisetup{table-format=1.4}
  \begin{tabular}{l S S c}
    \toprule
    Ion & {$\sigma_{ii}$ (nm)} & {$\epsilon_{ii}$ (kJ\,mol$^{-1}$)} & {$q_i$ ($e$)} \\
    \midrule
    Li$^+$ & 0.1285 & 2.525 & +1 \\
    Na$^+$ & 0.2310 & 0.450 & +1 \\
    Cl$^-$ & 0.4300 & 0.420 & -1 \\
    I$^-$  & 0.4730 & 1.380 & -1 \\
    \bottomrule
  \end{tabular}
\end{table}

\subsection{Convergence and relaxation of confined particle numbers}
\label{si:count_convergence}

\begin{figure}[H]
  \centering
  \includegraphics[width=\linewidth]{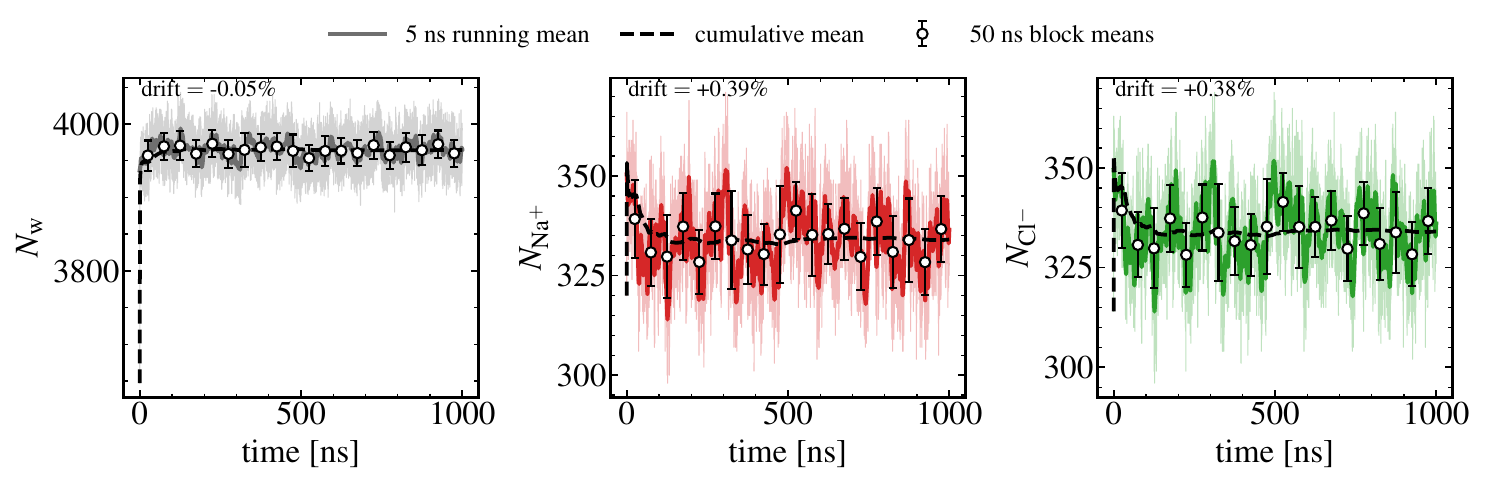}
  \caption[Convergence of confined particle numbers for NaCl]{Convergence of confined particle numbers for NaCl at
  \(H=\SI{4.08}{nm}\) and \(C_{\rm salt}^{\rm ave}=\SI{5}{M}\). From left to right, the panels show the numbers of confined water molecules \(N_w\), Na\(^+\) ions \(N_{\mathrm{Na}^+}\), and Cl\(^-\) ions \(N_{\mathrm{Cl}^-}\) in the central slit region over the \(\SI{1}{\micro\second}\) production trajectory. Particle numbers are sampled every \(\SI{20}{ps}\). Thin coloured traces show the raw particle numbers, solid coloured curves show \(\SI{5}{ns}\) running means, dashed black curves show cumulative means, and open circles with error bars show \(\SI{50}{ns}\) block means and standard deviations. The relative change between the first- and second-half means is given in each panel.}
  \label{fig:si_count_convergence}
\end{figure}

We first examine the time dependence of the numbers of water molecules, Na\(^+\) ions, and Cl\(^-\) ions in the central slit region to assess
the sampling of the confined composition. Figure~\ref{fig:si_count_convergence} shows a representative NaCl system at \(H=\SI{4.08}{nm}\) and \(C_{\rm salt}^{\rm ave}=\SI{5}{M}\) over the full \(\SI{1}{\micro\second}\) trajectory. The particle numbers fluctuate around stationary mean values. The relative changes between the first- and second-half means are \(-0.05\%\) for water, \(+0.39\%\) for Na\(^+\), and \(+0.38\%\) for Cl\(^-\), indicating stable confined water and ion populations over the production trajectory.

\begin{figure}[H]
  \centering
  \includegraphics[width=0.8\linewidth]{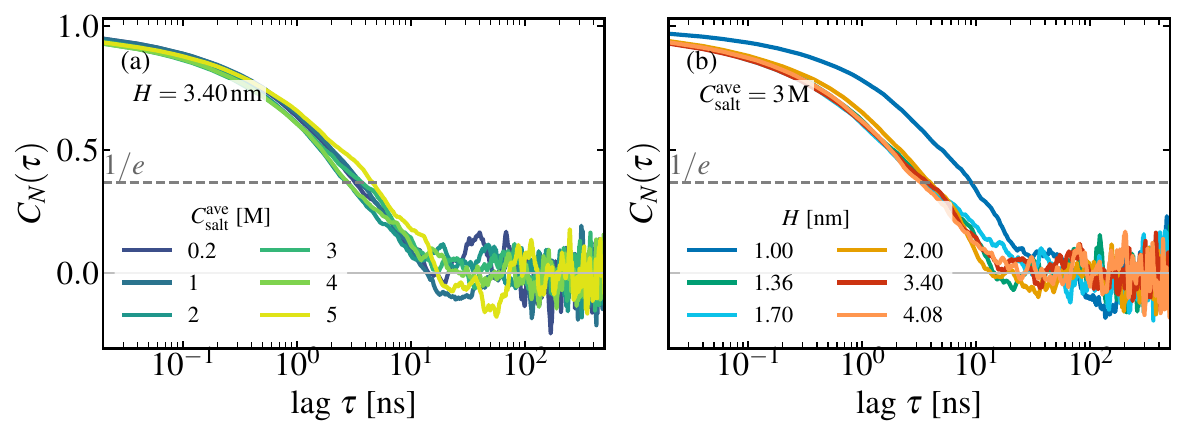}
  \caption[Autocorrelation of the confined salt number for NaCl]{Autocorrelation of the confined salt number for NaCl. The salt count is defined as \(N_{\rm salt}=\tfrac12(N_{\mathrm{Na}^+}+N_{\mathrm{Cl}^-})\), and \(C_N(\tau)\) denotes its normalised autocorrelation function (Eq.~\eqref{eq:si_salt_number_acf}).
  \textbf{(a)} \(C_N(\tau)\) for different \(C_{\rm salt}^{\rm ave}=0.2,\,1,\,2,\,3,\,4,\) and
  \(\SI{5}{M}\) at slit height \(H=\SI{3.40}{nm}\).
  \textbf{(b)} \(C_N(\tau)\) for
  \(H=1.00,\,1.36,\,1.70,\,2.00,\,3.40,\) and \(\SI{4.08}{nm}\) at \(C_{\rm salt}^{\rm ave}=\SI{3}{M}\). The horizontal dashed line marks \(1/e\), and the grey horizontal line marks zero correlation. The salt-number correlation time \(\tau_N\) is estimated by integrating \(C_N(\tau)\) from zero lag to its first zero crossing according to Eq.~\eqref{eq:si_salt_number_tau}.}
  \label{fig:si_salt_number_acf}
\end{figure}

To quantify the relaxation of fluctuations in the confined salt
population, we use the salt count
\(
  N_{\rm salt}(t)
  =
  \frac{1}{2}
  \left[
  N_{\mathrm{Na}^+}(t)+N_{\mathrm{Cl}^-}(t)
  \right]
\)
and its normalised autocorrelation function
\begin{equation}
  C_N(\tau)
  =
  \frac{
    \left\langle
    \delta N_{\rm salt}(t)
    \delta N_{\rm salt}(t+\tau)
    \right\rangle
  }{
    \left\langle
    \delta N_{\rm salt}^{\,2}
    \right\rangle
  },
  \qquad
  \delta N_{\rm salt}(t)
  =
  N_{\rm salt}(t)-\left\langle N_{\rm salt}\right\rangle .
  \label{eq:si_salt_number_acf}
\end{equation}
We estimate the salt-number correlation time from the integrated autocorrelation function truncated at its first zero crossing,
\begin{equation}
  \tau_N
  =
  \int_0^{\tau_0} C_N(\tau)\,d\tau ,
  \label{eq:si_salt_number_tau}
\end{equation}
where \(\tau_0\) is the first zero crossing of \(C_N(\tau)\). 

Figure~\ref{fig:si_salt_number_acf} shows that fluctuations of the confined salt population relax on a timescale of a few nanoseconds for most systems. At fixed \(H=\SI{3.40}{nm}\), \(\tau_N\) ranges from approximately \(\SI{3.5}{ns}\) to \(\SI{6.6}{ns}\) over \(C_{\rm salt}^{\rm ave}=0.2\)--\(\SI{5}{M}\). At fixed \(C_{\rm salt}^{\rm ave}=\SI{3}{M}\), the longest correlation time is found for the narrowest slit, \(H=\SI{1.00}{nm}\), for which \(\tau_N=\SI{10.84}{ns}\); for the wider slits,
\(\tau_N\approx\SIrange{3.7}{5.2}{ns}\). Thus, even for the system with the longest relaxation time, the \(\SI{1}{\micro\second}\) trajectory spans more than 90 salt-number correlation times.

\section{Damped-cosine fit of $\delta_w(H)$}
\label{si:dwfit}
The oscillatory approach of $\delta_w(H)$ to its wide-slit limit in Fig.~\ref{fig:pure_water}b is described with the empirical damped-cosine form in Eq.~\eqref{eq:damped_cosine}. The asymptotic value is fixed to the pure-water deficit length, $\delta_w^{\infty}=\SI{0.1919}{nm}$, while the remaining fitted parameters are $A_0=\SI{0.282}{nm}$, $\xi=\SI{0.366}{nm}$, $\lambda=\SI{0.325}{nm}$, and
$H_0=\SI{-0.472}{nm}$. The fit gives $R^2=0.921$ and $\mathrm{RMSE}=\SI{4.44e-3}{nm}$. Here, $A_0$ sets the amplitude of the finite-$H$ modulation, $\xi$ sets the decay length of the oscillations, $\lambda$ gives the oscillation period, and $H_0$ is the phase offset. The fitted period $\lambda\approx d_{\rm graphite}=\SI{0.335}{nm}$ is consistent with successive water-layer formation as the slit height increases.

\section{Water concentration response in electrolyte-filled slits}
\label{si:water_electrolyte}

\begin{figure}[H]
\centering
\includegraphics[width=\linewidth]{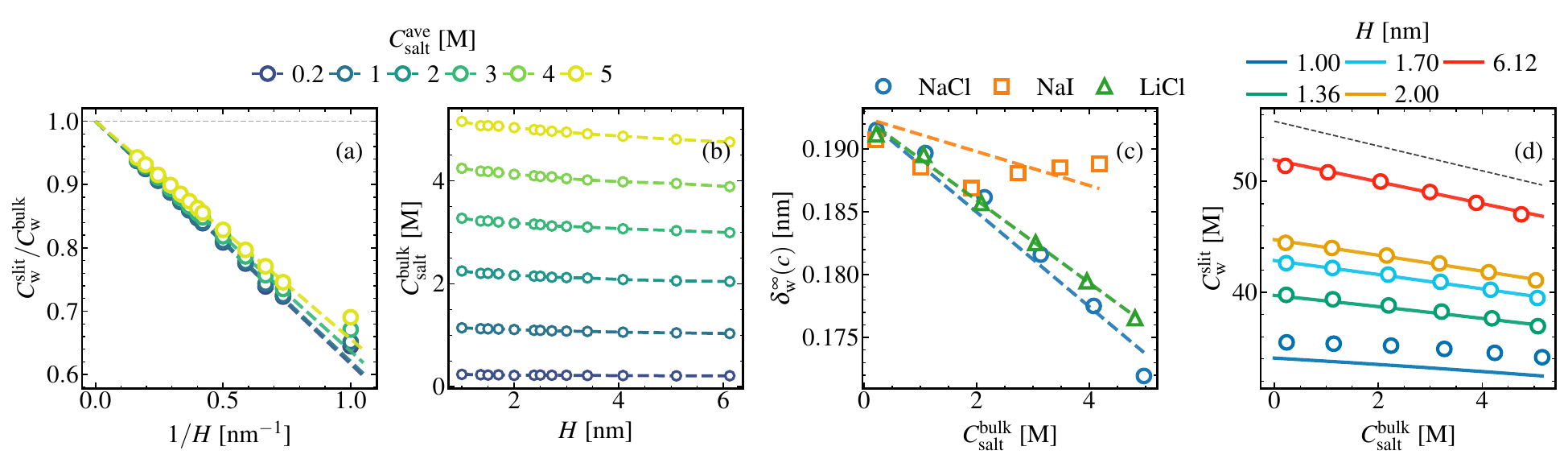}
\caption[Water concentration response in electrolyte-filled graphene slits]{Water concentration response in electrolyte-filled graphene slits.
\textbf{(a)} Slit-to-bulk water concentration ratio \(C_w^{\rm slit}/C_w^{\rm bulk}\) for NaCl as a function of \(1/H\) for different system-wide average salt concentrations \(C_{\rm salt}^{\rm ave}\). Dashed lines are fixed-intercept fits of Eq.~\eqref{eq:water_ratio}, with the intercept constrained to unity in the infinite-height limit, and are used to extract \(\delta_w^{\infty}\) for each \(C_{\rm salt}^{\rm ave}\) series.
\textbf{(b)}Reservoir salt concentration
\(C_{\rm salt}^{\rm bulk}\) as a function of slit height \(H\) for different \(C_{\rm salt}^{\rm ave}\).
\textbf{(c)} Concentration-dependent water deficit length
\(\delta_w^{\infty}(C_{\rm salt}^{\rm bulk})\) for NaCl, NaI, and LiCl. Dashed lines are linear fits constrained to the pure-water value \(\delta_w^{\infty}(C_{\rm salt}^{\rm bulk}=0) =\SI{0.1919}{nm}\), obtained from Fig.~\ref{fig:pure_water}b. The NaCl fit is used in panel~(d).
\textbf{(d)} Slit water molarity \(C_w^{\rm slit}\) for NaCl as a function of \(C_{\rm salt}^{\rm bulk}\) for different slit heights \(H\). Symbols are simulation data and solid lines are predictions from Eq.~\eqref{eq:water_pred}, using
\(\delta_w^{\infty}(C_{\rm salt}^{\rm bulk})\) from panel~(c) and a linear fit to the reservoir water molarity
\(C_w^{\rm bulk}(C_{\rm salt}^{\rm bulk})\). The black dashed line shows the fitted reservoir water molarity from Eq.~\eqref{eq:water_bulk_fit}. All concentrations are averaged over the central slit and reservoir plateau windows defined in Fig.~\ref{fig:system}.}
\label{fig:water_electrolyte}
\end{figure}

The main text focuses on salt depletion in electrolyte solutions. The
corresponding water response is shown here. At fixed
\(C_{\rm salt}^{\rm ave}\), the water ratio
\(C_w^{\rm slit}/C_w^{\rm bulk}\) decreases approximately linearly
with \(1/H\). This follows the same deficit-length form used in the
main text,
\begin{equation}
C_w^{\rm slit}/C_w^{\rm bulk}
=
1 - 2\delta_w^{\infty}/H,
\label{eq:water_ratio}
\end{equation}
as shown by the dashed lines for NaCl in Fig.~\ref{fig:water_electrolyte}a. The intercept is fixed to unity in the infinite-height limit, and the slope \(-2\delta_w^{\infty}\) yields the deficit length for each system-wide average salt concentration \(C_{\rm salt}^{\rm ave}\). 

At fixed \(C_{\rm salt}^{\rm ave}\), \(C_{\rm salt}^{\rm bulk}\) varies slightly with \(H\), as shown in Fig.~\ref{fig:water_electrolyte}b. Each extracted deficit length represents one \(C_{\rm salt}^{\rm ave}\) series, so we plot it at the mean reservoir concentration of that series. This applies to the deficit-length plots in Fig.~\ref{fig:nacl_summary}c, Fig.~\ref{fig:water_electrolyte}c, and Fig.~\ref{fig:transport_nacl}d. For NaCl these mean concentrations are \(0.222\), \(1.091\), \(2.141\), \(3.140\), \(4.079\), and \(\SI{4.970}{M}\), for \(C_{\rm salt}^{\rm ave}=0.2,\,1,\,2,\,3,\,4,\) and
\(\SI{5}{M}\). In all other plots where \(C_{\rm salt}^{\rm bulk}\) is used for data at individual slit heights, we use the reservoir concentration of each slit state.

Figure~\ref{fig:water_electrolyte}c shows the resulting
\(\delta_w^{\infty}(C_{\rm salt}^{\rm bulk})\) values for NaCl, NaI, and LiCl from Eq.~\eqref{eq:water_ratio}. To describe their dependence on reservoir salt concentration, we fit these values linearly as functions of \(C_{\rm salt}^{\rm bulk}\), as shown by the dashed lines in Fig.~\ref{fig:water_electrolyte}c. The linear fits are constrained to the pure-water value \(\delta_w^{\infty}(C_{\rm salt}^{\rm bulk}=0)
=\SI{0.1919}{nm}\), obtained from Fig.~\ref{fig:pure_water}b. For NaCl, this fit provides the
concentration-dependent deficit length used in Fig.~\ref{fig:water_electrolyte}d.

The slit water molarity also depends on the reservoir water molarity,
which decreases as the salt concentration increases. For NaCl, a linear
fit to the reservoir values gives
\begin{equation}
C_w^{\rm bulk}(C_{\rm salt}^{\rm bulk})
= -1.1161\,C_{\rm salt}^{\rm bulk} + \SI{55.4273}{M},
\label{eq:water_bulk_fit}
\end{equation}
with \(R^2=0.991\) over
\(C_{\rm salt}^{\rm bulk}=\SIrange{0.21}{5.15}{M}\). This fit is shown
by the black dashed line in Fig.~\ref{fig:water_electrolyte}d.

Combining the fitted reservoir water molarity with the NaCl deficit
length gives
\begin{equation}
C_w^{\rm slit}(C_{\rm salt}^{\rm bulk},H)
\approx C_w^{\rm bulk}(C_{\rm salt}^{\rm bulk})
\left[ 1 - \frac{2\delta_w^{\infty}(C_{\rm salt}^{\rm bulk})}{H} \right].
\label{eq:water_pred}
\end{equation}
The resulting predictions are shown by the solid lines in
Fig.~\ref{fig:water_electrolyte}d. They capture the dependence on slit height and salt concentration, with visible deviations only for \(H=\SI{1.00}{nm}\), where water forms a pronounced two-layer structure.

\section{Bulk-reference and slit-only transport simulations}
\label{si:slitonly}

\subsection{Slit-only simulations for transport}
To quantify the transport properties of the nanoslit, we perform slit-only simulations in a periodic cell with
\(L_y=\SI{8.85}{nm}\) and \(L_z=\SI{8.52}{nm}\). The number of particles in each slit-only cell is obtained by multiplying the mean number of particles in the central slit window of the corresponding slit--reservoir run by \(L_z/L_c\) ( \(L_c=4\) nm), the ratio of the slit-only cell length to the central-window length. This scaling preserves the mean confined-fluid concentration. These cells are energy-minimised, equilibrated first in the NPT ensemble, and subsequently equilibrated in the NVT ensemble using the force field and settings of Sec.~\ref{si:md}.

The slit-only Green--Kubo self-diffusivities and ionic conductivity are obtained from \(\SI{10}{ns}\) NVT production runs using a \(\SI{2}{fs}\) time step, with velocities stored every other step (\(\SI{4}{fs}\)). The total centre-of-mass velocity is removed every 100 steps. Separate finite-field slit-only simulations are run for \(\SI{300}{ns}\) with fields of \(0.05\) and \(\SI{0.10}{V\,nm^{-1}}\) along the channel. Coordinates and velocities are stored every \(\SI{1}{ps}\), and the steady-state current is averaged over the final \(\SI{280}{ns}\). The details for the analysis performed to extract the transport quantities are given below.

\subsection{Bulk reference simulations for transport}
Independent bulk-reference transport properties are obtained from fully periodic cubic cells containing pure water or NaCl solutions at initial salt concentrations of \(0.2\), \(1\), \(2\), \(3\), \(4\), and \(5\) M. The salt cells contain, respectively, \(2950/11\), \(2862/55\), \(2752/110\), \(2642/165\), \(2532/220\), and \(2424/274\) water molecules/NaCl ion pairs. After energy minimisation, each system is equilibrated for \(\SI{0.5}{ns}\) in the NVT ensemble at \(\SI{300}{K}\), followed by \(\SI{1}{ns}\) of isotropic NPT equilibration at \(\SI{300}{K}\) and \(\SI{1}{bar}\). The resulting cubic box lengths range from \(\SI{4.48}{nm}\) at \(0.2\) M to \(\SI{4.35}{nm}\) at \(5\) M. The corresponding salt concentrations calculated from the ion-pair numbers and equilibrated box volumes are \(0.204\), \(1.033\), \(2.107\), \(3.237\), \(4.392\), and \(\SI{5.511}{M}\), respectively. Because the \(C_{\rm salt}^{\rm bulk}\) of each slit state
does not exactly match one of these bulk-reference concentrations, we
obtain the corresponding bulk self-diffusivity and Green--Kubo
conductivity by linear interpolation between the two neighbouring
bulk-reference simulations. These interpolated values are used in the
slit-to-bulk transport ratios in Fig.~\ref{fig:transport_nacl}c.

Bulk production simulations are then performed for \(\SI{10}{ns}\) in the NVT ensemble using a \(\SI{2}{fs}\) time step. Velocities are stored every 10 steps (\(\SI{20}{fs}\)) for the velocity- and charge-current autocorrelation functions used to obtain the bulk self-diffusivity and Green--Kubo conductivity. The total centre-of-mass velocity is removed every 100 steps.

\subsection{Velocity autocorrelation and Green--Kubo self-diffusion}
\label{si:vacf}

\begin{figure*}[!t]
  \centering
  \includegraphics[width=0.65\linewidth]{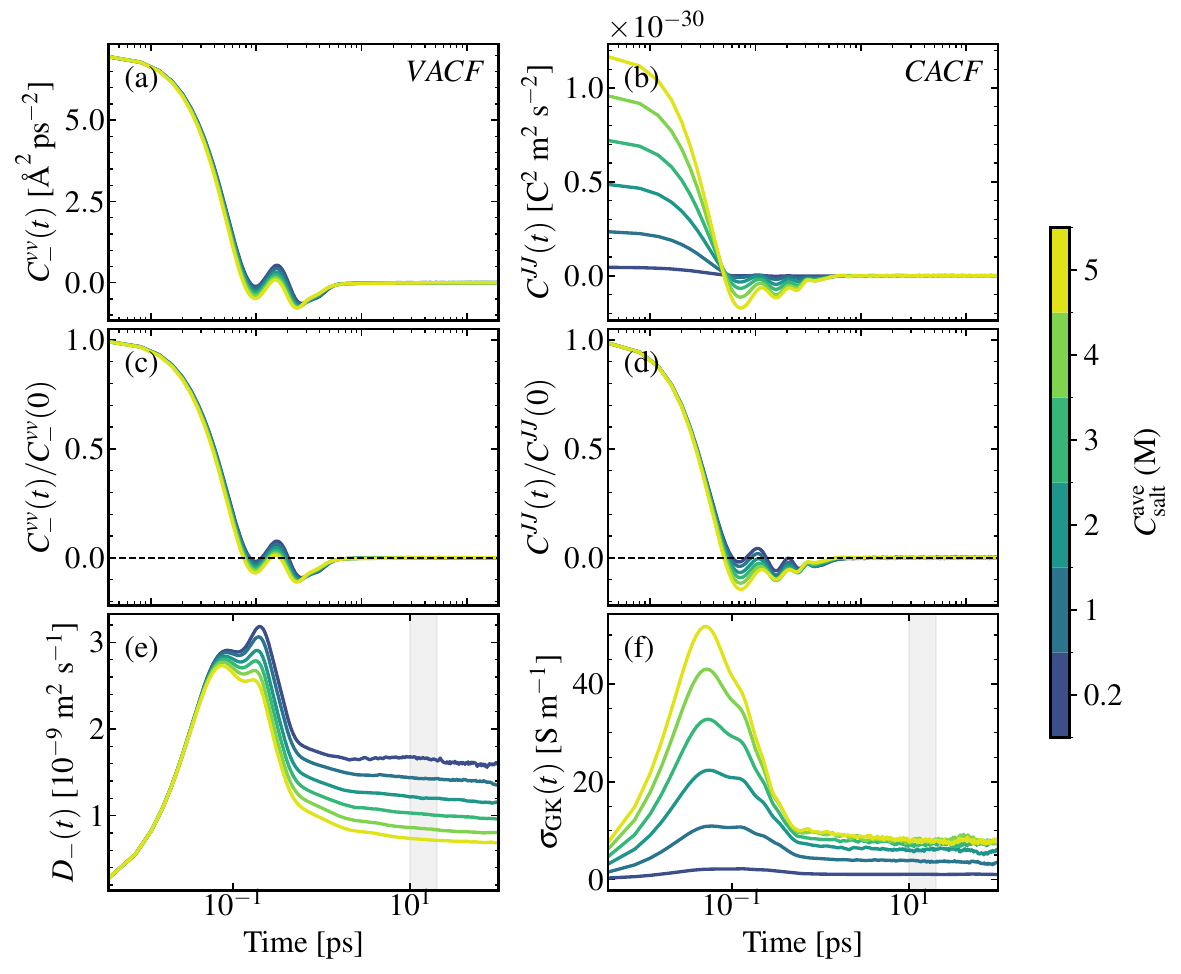}
  \caption[Green--Kubo extraction of anion self-diffusion and ionic conductivity]{Green--Kubo extraction of anion self-diffusion and ionic conductivity, illustrated for NaCl in the \(H=\SI{2.00}{nm}\) slit. The reported in-plane quantities are averages over the two directions parallel to the graphene sheets, \(y\) and \(z\). Curves are coloured by system-wide average salt concentration \(C_{\rm salt}^{\rm ave}\) (\(0.2\)--\(\SI{5}{M}\)).
  \textbf{(a)} In-plane anion velocity autocorrelation function (VACF)
  \(C_-^{vv}(t)=\tfrac12[C_-^{vv,y}(t)+C_-^{vv,z}(t)]\).
  \textbf{(b)} In-plane charge-current autocorrelation function (CACF)
  \(C^{JJ}(t)=\tfrac12[C_y^{JJ}(t)+C_z^{JJ}(t)]\).
  \textbf{(c,d)} The corresponding VACF and CACF normalised by their \(t=0\) values.
  \textbf{(e)} Running integral of the in-plane anion VACF, giving \(D_-(t)\).
  \textbf{(f)} Running integral of the in-plane CACF, giving
  \(\sigma_{\rm GK}(t)\). The shaded band in (e,f) marks the plateau window \([t_1,t_2]\) used to define \(D_-\) and
  \(\sigma_{\rm GK}\).}
  \label{fig:si_vacf_cacf_triptych}
\end{figure*}

For species \(X\) and Cartesian component \(\alpha\), the velocity autocorrelation function (VACF) is
\begin{equation}
  C_X^{vv,\alpha}(t)
  =
  \frac{1}{n_X}
  \sum_{i=1}^{n_X}
  \left\langle
  v_i^{\alpha}(0)v_i^{\alpha}(t)
  \right\rangle ,
  \label{eq:si_vacf}
\end{equation}
where \(n_X\) is the number of particles of species \(X\) in the slit-only system and \(v_i^{\alpha}\) is the \(\alpha\)-component of the velocity of particle \(i\). For the slit geometry, the in-plane VACF is the average over the two directions parallel to the graphene sheets,
\begin{equation}
  C_X^{vv}(t)
  =
  \frac{1}{2}
  \left[
  C_X^{vv,y}(t)+C_X^{vv,z}(t)
  \right].
  \label{eq:si_vacf_inplane}
\end{equation}
The corresponding running Green--Kubo integral is
\begin{equation}
  D_X(t)
  =
  \int_0^t C_X^{vv}(\tau)\,d\tau ,
  \label{eq:si_vacf_running}
\end{equation}
and the reported in-plane self-diffusion \(D_X\) is
obtained by averaging \(D_X(t)\) over the first stable plateau window.

Figure~\ref{fig:si_vacf_cacf_triptych}(a,c,e) shows the representative anion VACF, its normalised form, and the running integral \(D_-(t)\) for the \(H=\SI{2.00}{nm}\) slit at all sampled \(C_{\rm salt}^{\rm ave}\). The normalised VACFs have similar decay shapes across concentrations, and the running integrals reach stable plateaus by \(t\approx\SI{10}{ps}\), which are used to extract \(D_-\).

\subsection{Charge-current autocorrelation and Green--Kubo conductivity}
\label{si:cacf}
For ion \(i\), the charge is \(Q_i=q_i e\), where \(q_i\) is the charge
in units of the elementary charge \(e\). The ionic charge current is
\begin{equation}
  \mathbf{J}(t)
  =
  \sum_{i\in\mathrm{ions}} Q_i\mathbf{v}_i(t).
  \label{eq:si_current}
\end{equation}
Fixed carbons and water molecules do not contribute to the net ionic
current. For Cartesian component \(\alpha\), the charge-current autocorrelation function (CACF) is
\begin{equation}
  C_\alpha^{JJ}(t)
  =
  \left\langle J_\alpha(0)J_\alpha(t)\right\rangle .
\end{equation}
For the slit geometry, the in-plane CACF is the average over the two directions parallel to the graphene sheets,
\begin{equation}
  C^{JJ}(t)
  =
  \frac{1}{2}
  \left[
  C_y^{JJ}(t)+C_z^{JJ}(t)
  \right].
  \label{eq:si_cacf_inplane}
\end{equation}
The corresponding running Green--Kubo integral is
\begin{equation}
  \sigma_{\rm GK}(t)
  =
  \frac{1}{V k_B T}
  \int_0^t C^{JJ}(\tau)\,d\tau ,
  \label{eq:si_cacf_running}
\end{equation}
where \(V=HL_yL_z\) is the slit-only system volume, \(k_B\) is Boltzmann's constant, and \(T\) is the temperature. The reported in-plane Green--Kubo conductivity \(\sigma_{\rm GK}\) is obtained by averaging \(\sigma_{\rm GK}(t)\) over the first stable plateau window.

Figure~\ref{fig:si_vacf_cacf_triptych}(b,d,f) shows the representative CACF, its normalised form, and the running integral \(\sigma_{\rm GK}(t)\) for the same \(H=\SI{2.00}{nm}\) slit. The absolute CACF amplitude increases strongly with salt concentration, reflecting the larger number of charge carriers. After normalisation, the decay shapes are much more similar across concentrations, and the running integrals reach stable plateaus on the same \(\sim\SI{10}{ps}\) timescale as \(D_-(t)\).

\subsection{Finite-field conductivity}
\label{si:ff}

\begin{figure*}[!t]
  \centering
  \includegraphics[width=0.9\linewidth]{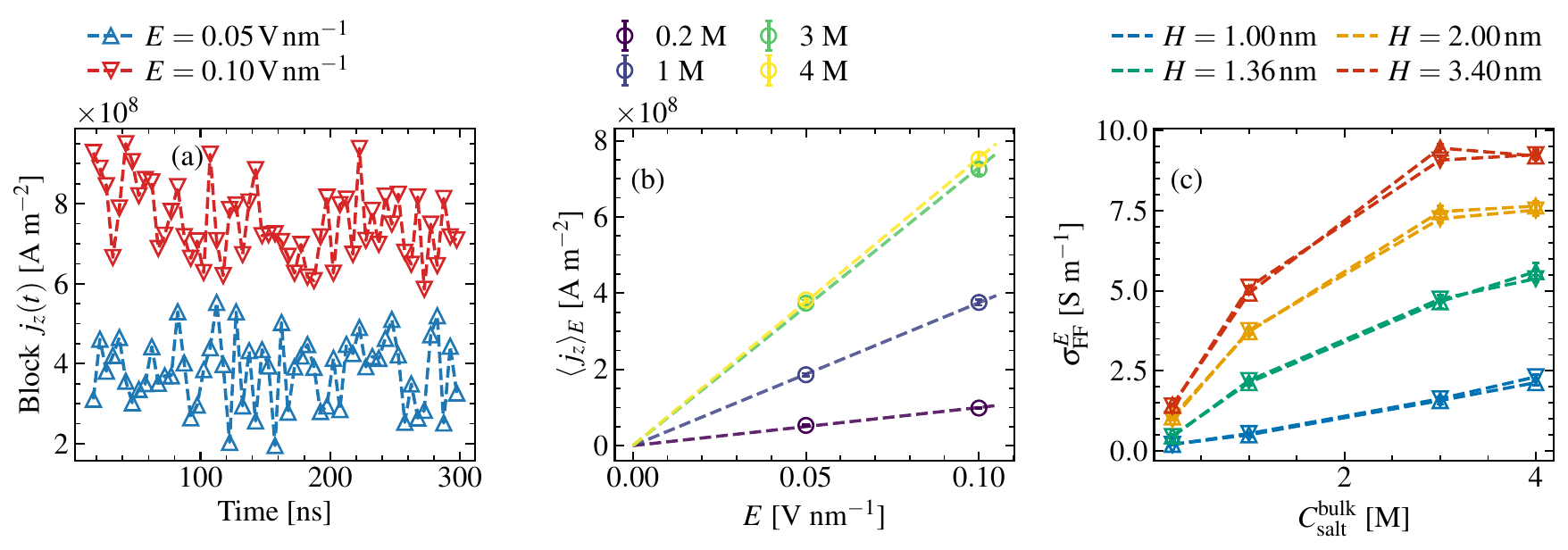}
  \caption[Finite-field conductivity validation for NaCl slit-only simulations]{Finite-field conductivity validation for NaCl slit-only simulations.
  \textbf{(a)} Block-averaged in-plane current density \(j(t)\) for the representative slit \(H=\SI{2.00}{nm}\) at \(C_{\rm salt}^{\rm ave}=\SI{3.0}{M}\), shown for  \(E=\SI{0.05}{V/nm}\) and \(\SI{0.10}{V/nm}\).
  \textbf{(b)} Mean current density \(\langle j\rangle_E\) versus applied field for \(H=\SI{2.00}{nm}\) at the reservoir salt concentrations shown in the legend; dashed lines are linear fits through the origin.
  \textbf{(c)} Finite-field conductivity \(\sigma_{\rm FF}^{E}=\langle j\rangle_E/E\) as a function
  of \(C_{\rm salt}^{\rm bulk}\) for the slit heights shown by colour. Symbols distinguish \(\sigma_{\rm FF}^{0.05}\) and \(\sigma_{\rm FF}^{0.10}\). The similar conductivities obtained from the two applied fields indicate that the finite-field simulations remain close to the linear-response regime over the field range used here.}
  \label{fig:si_ff_validation}
\end{figure*}

Finite-field conductivities are obtained from additional NVT
simulations with a uniform electric field applied along the slit \(z\) direction,
\(
  E\in\{0.05,\,0.10\}\ \si{V/nm}.
\)
The instantaneous ionic current density along the applied field is
\begin{equation}
  j(t)
  =
  \frac{1}{V}
  \sum_{i\in\mathrm{ions}} Q_i v_i(t),
  \label{eq:si_ff_current_density}
\end{equation}
where \(v_i(t)\) is the velocity component along the applied field and
\(V\) is the slit-only system volume. After discarding the initial
transient, the field-specific finite-field conductivity is obtained from
the steady-state mean current density,
\begin{equation}
  \sigma_{\rm FF}^{E}
  =
  \frac{\langle j\rangle_E}{E}.
  \label{eq:si_ff_sigma}
\end{equation}

Figure~\ref{fig:si_ff_validation} shows the finite-field validation for NaCl slit-only systems. Panels~(a,b) use the \(H=\SI{2.00}{nm}\) slit as an example: the block-averaged current density remains stable over the production window, and the mean current density increases approximately linearly with the applied field. Panel~(c) summarises the two field-specific conductivities, \(\sigma_{\rm FF}^{0.05}\) and \(\sigma_{\rm FF}^{0.10}\), for the slit heights used in the main text. The two applied fields give similar conductivities over the sampled concentrations and slit heights, supporting their use as finite-field estimates of the linear-response
conductivity.

\subsection{Self-diffusion of ions and water}
\begin{figure*}[!t]
  \centering
  \includegraphics[width=0.7\linewidth]{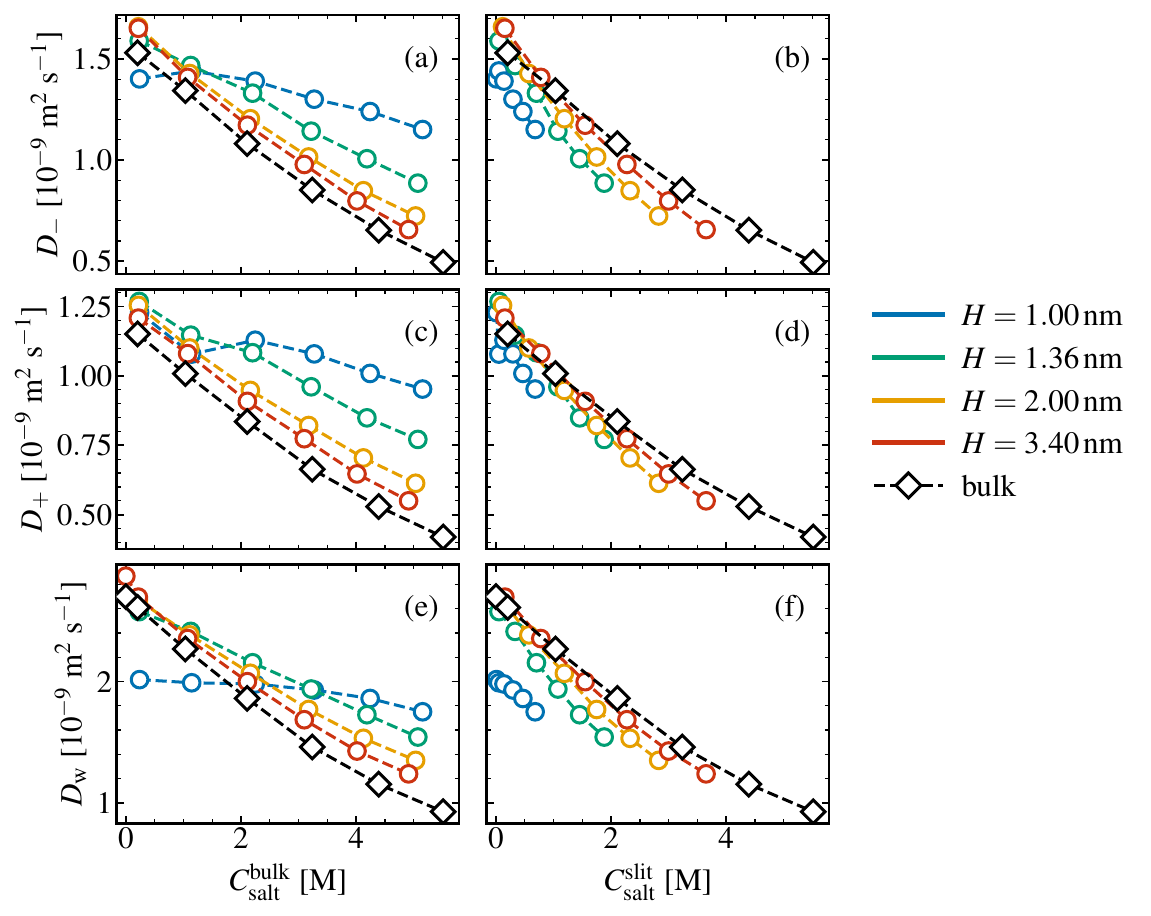}
  \caption[Self-diffusion for NaCl in the slit-only geometry]{Self-diffusion for NaCl in the slit-only
  systems, plotted against reservoir and slit salt concentration.
  \textbf{(a,c,e)} \(D_-\), \(D_+\), and \(D_w\) as functions of \(C_{\rm salt}^{\rm bulk}\) for slit heights \(H=1.00,\,1.36,\,2.00,\,3.40\)~nm, together with the bulk-reference data (black dashed line).
  \textbf{(b,d,f)} The same data replotted against the slit salt
  concentration \(C_{\rm salt}^{\rm slit}\).}
  \label{fig:si_selfD_cbulk_cslit}
\end{figure*}

Figure~\ref{fig:si_selfD_cbulk_cslit} summarises the plateau-extracted self-diffusion for the three mobile species \(X\in\{-,+,w\}\) across all sampled slit heights and salt loadings. When plotted against \(C_{\rm salt}^{\rm bulk}\), the slit curves lie above the bulk reference. Replotting against \(C_{\rm salt}^{\rm slit}\) reduces this concentration offset.

\section{Concentration-dependent deficit lengths}
\label{si:deltaC}
The coefficients in Eq.~\eqref{eq:deltaC} are obtained by fitting the NaCl \(\delta_\Theta^{\infty}\) values as a function of the mean \(C_{\rm salt}^{\rm bulk}\) for each \(C_{\rm salt}^{\rm ave}\) series. Linear fits are used for all observables over the sampled concentration range:
\begin{align}
  \delta_{w}^{\infty}(C_{\rm salt}^{\rm bulk})
    &= 0.1919 - 0.0036\,C_{\rm salt}^{\rm bulk},\\
  \delta_{{\rm salt}}^{\infty}(C_{\rm salt}^{\rm bulk})
    &= 0.4898 - 0.0123\,C_{\rm salt}^{\rm bulk},\\
  \delta_{\sigma}^{\infty}(C_{\rm salt}^{\rm bulk})
    &= 0.5260 - 0.0437\,C_{\rm salt}^{\rm bulk},\\
  \delta_{D_-}^{\infty}(C_{\rm salt}^{\rm bulk})
    &= -0.0238 - 0.0627\,C_{\rm salt}^{\rm bulk},\\
  \delta_{D_+}^{\infty}(C_{\rm salt}^{\rm bulk})
    &= -0.0494 - 0.0680\,C_{\rm salt}^{\rm bulk},\\
  \delta_{D_w}^{\infty}(C_{\rm salt}^{\rm bulk})
    &= 0.0149 - 0.0707\,C_{\rm salt}^{\rm bulk}.
\end{align}

Here \(C_{\rm salt}^{\rm bulk}\) is in M and the deficit lengths are in nm. For the water concentration, the intercept is constrained to the pure-water deficit length,
\(
\delta_w^{\infty}(C_{\rm salt}^{\rm bulk}=0)
=\SI{0.1919}{nm}
\)
obtained from Fig.~\ref{fig:pure_water}b of the main text. All other intercepts are fitted freely. Table~\ref{tab:delta_coeffs} collects the same coefficients in the
generic \(\{\alpha_\Theta,\beta_\Theta\}\) form of Eq.~\eqref{eq:deltaC}.

\begin{table}[h]
  \centering
  \caption[Coefficients of the deficit-length fit for NaCl]{Coefficients of Eq.~\eqref{eq:deltaC} for NaCl.}
  \label{tab:delta_coeffs}
  \sisetup{table-format=+1.4}
  \begin{tabular}{l S S}
    \toprule
    \(\Theta\) &
    {\(\alpha_\Theta\) (nm)} &
    {\(\beta_\Theta\) (nm\,M\(^{-1}\))} \\
    \midrule
    \(C_w\)              & +0.1919 & -0.0036 \\
    \(C_{\rm salt}\)     & +0.4898 & -0.0123 \\
    \(\sigma\)  & +0.5260 & -0.0437 \\
    \(D_-\)              & -0.0238 & -0.0627 \\
    \(D_+\)              & -0.0494 & -0.0680 \\
    \(D_w\)              & +0.0149 & -0.0707 \\
    \bottomrule
  \end{tabular}
\end{table}

\begin{figure}[H]
  \centering
  \includegraphics[width=\linewidth]{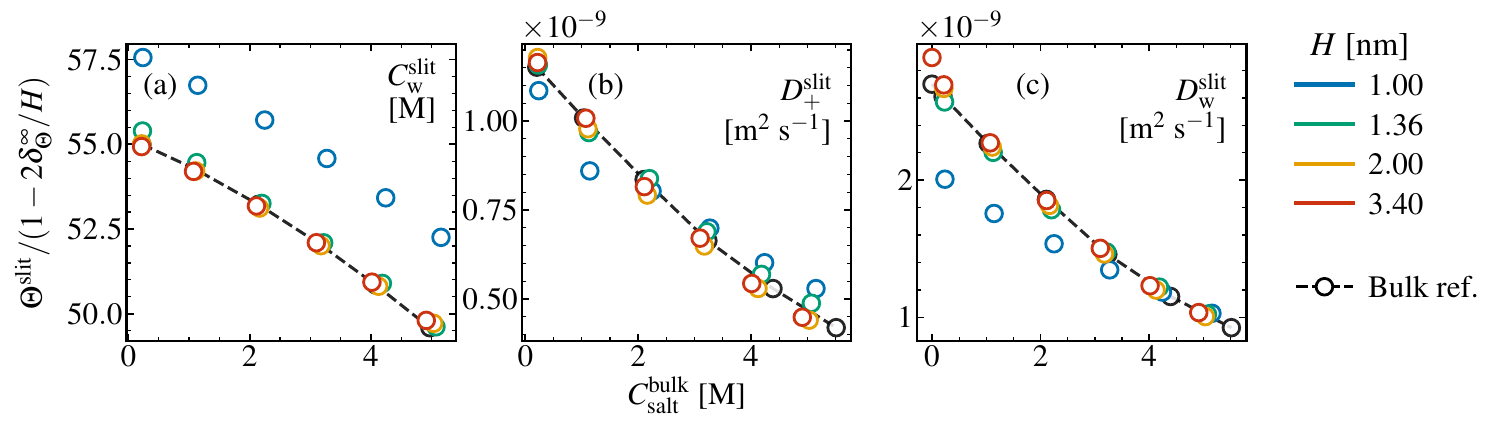}
  \caption[Deficit-length rescaling for the water and cation observables]{Deficit-length rescaling for the water and
  cation observables.
  \textbf{(a)} Water molarity \(C_w^{\rm slit}\).
  \textbf{(b)} Cation self-diffusion \(D_+^{\rm slit}\).
  \textbf{(c)} Water self-diffusion \(D_w^{\rm slit}\).
  The plotted quantity is
  \(\Theta^{\rm slit}/(1-2\delta_\Theta^{\infty}/H)\), using
  \(\delta_\Theta^{\infty}(C_{\rm salt}^{\rm bulk})\) from
  Eq.~\eqref{eq:deltaC}. Dashed black lines show the corresponding
  bulk references, and colours indicate different slit heights \(H\).}
  \label{fig:transport_corrected_si}
\end{figure}

Figure~\ref{fig:transport_corrected_si} shows the rescaled forms of the water molarity \(C_w^{\rm slit}\), cation self-diffusion \(D_+^{\rm slit}\), and water self-diffusion \(D_w^{\rm slit}\). These quantities complement the salt molarity, anion diffusion, and ionic conductivity shown in the main text. For \(H\ge\SI{1.36}{nm}\), the rescaled data closely follow their bulk references, indicating that their slit-height dependence is largely removed by the deficit-length rescaling. Deviations are only observed for \(H=\SI{1.00}{nm}\) slit: \(C_w^{\rm slit}\) remains above the
bulk reference after rescaling, and smaller residual offsets are also visible in \(D_+^{\rm slit}\) and \(D_w^{\rm slit}\). This behaviour is consistent with the strongly structured two-water-layer state at \(H=\SI{1.00}{nm}\), where the two interfacial regions are not yet separated by a bulk-like region.

\FloatBarrier

\bibliographystyle{unsrtnat}
\bibliography{ion_confinement}

\end{document}